\documentclass[aps,pra,preprint,showpacs]{revtex4-1}
\usepackage{amsmath,graphicx}
\usepackage{epsfig}
\usepackage{color}
\begin{document}
\def\la{{\langle}}
\def\ra{{\rangle}}
\def\a{{\alpha}}
\def\ah{\hat{A}}
\def\h{\hat{H}}
\def\q{{\quad}}

\title{'Superluminal paradox' in wavepacket propagation and its quantum mechanical resolution}
%
\author {D. Sokolovski$^{1,2}$ and E. Akhmatskaya$^3$}
\affiliation{$^1$ Department of Physical Chemistry,  University of the Basque Country, Leioa, Bizkaia, Spain,\\
$^2$ IKERBASQUE, Basque Foundation for Science, 48011, Bilbao, Spain,\\
$^3$Basque Center for Applied Mathematics (BCAM),\\ Alameda de Mazarredo, 14  48009 Bilbao Bizkaia, Spain }
 \date{\today}
 \email {Corresponding author: d.sokolovski@ehu.es or dgsokol15@gmail.com}
\begin{abstract}
We analyse in detail the reshaping mechanism leading to apparently 'superluminal' advancement of a wave packet traversing a classically forbidden region. 
In the coordinate representation, a barrier is shown to act as an effective beamsplitter, recombining envelopes of the freely propagating pulse with various spacial shifts. Causality ensures that none of the constituent envelopes are advanced with respect to free propagation, yet the resulting pulse is advanced due to a peculiar interference effect, similar to the one responsible for 'anomalous' values which occur in Aharonov's 'weak measurements'.
In the momentum space, the effect is understood as a bandwidth phenomenon, where the incident pulse probes local, rather than global, analytical properties of the transmission amplitude $T(p)$. 
The advancement is achieved when $T(p)$ mimics locally an exponential behaviour, similar to the one occurring in Berry's 'superoscillations'.
Seen in a broader quantum mechanical context, the 'paradox'  is but a consequence of an attempt to obtain 'which way?' information without destroying the interference between the pathways of interest. This explains, to a large extent, the failure to adequately describe tunnelling in terms of a single 'tunnelling time'. 

\end{abstract}

%
%
\maketitle
\section {Introduction}
In the early 1930's MacColl \cite{MCOLL} noticed that quantum tunnelling appears to take no time or little time, in the sense that a
wave packet,
transmitted across a classically forbidden region, may arrive at a detector earlier than the one that moves in free space. If the advanced peak is used to predict the time $\tau$ the particle has spent in the barrier region, the result is nearly zero. Dividing the barrier width by $\tau$ yields a velocity exceeding the speed of light $c$, suggesting further that the transmission has a 'superluminal' aspect. 
The effect has been predicted and observed for various systems such as potential barriers, semi-transparent mirrors, refraction of light, microwaves in undersized wave guides and fast-light materials (for a review see Refs. \cite{REV}-\cite{REV3}). 
\newline
Superluminal velocities are strictly forbidden by Einstein's causality, yet one might entertain the suspicion that it might be violated for extremely rare classically forbidden events.
Such radical interpretations can be found, for example, in Ref. \cite{NIM1} which states
that 'the effect ... violates relativistic causality', in \cite{NIM2} which suggests that 'evanescent waves do exist in a space free of time', and in Refs.\cite{NIM3}-\cite{NIM5} which report 'macroscopic violation of general relativity'.
\newline
There is, however, some consensus that there is no conflict with relativity, since the transmitted wave packet undergoes in the barrier region a drastic reshaping, \cite{RESH}-\cite{BUTT1}. As a result, the transmitted pulse is formed from the front part of the incident one.
\newline
An explanation incompatible with the views of Ref. \cite{BUTT1}, was proposed in Refs. \cite{REV3}, \cite{WIN1}-\cite{WIN3}. There it was argued that, as no propagation occurs in the classically forbidden region, the effect arises from the energy storage in the barrier region. One consequence of this argument is that the transmitted wave packet (pulse) should not be 'frontloaded', i.e., should not arise from the front of the incident pulse. 
To our knowledge, the issue has not been fully resolved to date. 
\newline
There have been also other approaches to understanding the 'speed up effect' observed in tunnelling, of which we name here a few. A method for classifying various time parameters constructed to describe the processes of transmission and reflection was proposed in \cite{M1}. In Ref. \cite{M2} the authors used the influence functional technique to analyse the correlation between the initial and final positions of the transmitted particles. In the Bohm's formulation of quantum mechanics it was demonstrated that the transmitted wave packet builds up from Bohm's trajectories emanating from the front tail of the incident one \cite{L1}-\cite{L3}. A review of an alternative approach based on the analysis of the time at which a scattered particle arrives at a given spatial location can be found in Ref. \cite{M3}. Finally, a detailed analysis of the transient effects in propagation of matter waves was given in \cite{M4}.
\newline
The main purpose of this paper
is to provide a detailed analysis of the reshaping mechanism leading to the 'superluminal paradox' just described.
Various aspects of the problem have been studied in detail in Refs. \cite{DS1}-\cite{DS4a}, 
to which the reader is referred for mathematical justification.
We will confirm that the apparent 'superluminality' is an essentially interference effect, whose mechanism cannot be reduced to a naive reshaping or the energy storage argument.
Moreover, it occurs due to a particular type of interference known in 
quantum mechanics, or more precisely, in the quantum measurement theory.
It was shown in Ref.\cite{DS1} that the reshaping of the transmitted pulse occurs through an interference effect very similar to the one which causes the appearance of 'anomalous' values in the so-called 'weak' quantum measurement introduced by Aharonov and co-workers in  \cite{AH1}-\cite{AH3}.  Using this analogy, we will demonstrate that the measured quantity is the spacial shift of the transmitted particle, a variable 'conjugate' to the particle's momentum, which bears no direct relation to the duration of a tunnelling event \cite{FOOTSC}. We argue that our effort is justified by the benefit of bringing the problem into the framework of conventional quantum theory, 
and explicitly avoiding the notion of unduly short 'tunnelling times' and illegal 'superluminal velocities'.
\newline
The rest of the paper is organised as follows. In Sect.II we briefly discuss a naive view of how  reshaping may occur. Sections III - X contain our analysis of a quantum reshaping mechanism in a model which,
despite its simplicity, captures the main features of the 'superluminal' effect in tunnelling. In Sections XI - XVI we extend the analysis to the case of tunnelling across a classically forbidden region, or a wave propagation in an undersized waveguide. In Sect. XVII we briefly compare our approach to the one based on evaluating time variation of the signal at a fixed location \cite{JAPHA}. In Section XVIII we discuss the role of other quantum time parameters such as 
the traversal (Larmor) time.
Section XIX is a brief review of the physical explanations of the effect proposed earlier in the literature.
Section XX contains our concluding remarks.
\section {A naive view of reshaping}
As the name suggests, reshaping implies change of an object's shape in a way that some reference point, e.g., a peak, the centre of mass, or an edge is moved to a different location.
In the literature there are many similar examples of reshaping (see, e.g.,  \cite{RESH}, \cite{WIN3}), and here we risk one of our own. Suppose one sends 'signals' in the form of unilateral triangular shapes moving from left to right at a constant velocity $v_0$. A signal is 'received' when its peak passes through a certain remote point where the detector is placed.  Suppose that one of two identical 'signals' passes behind a screen of  a width $L$. There, someone, unseen by the observer, uses scissors (the triangles are made of paper) to cut  a similar yet much smaller triangle from the front part of the shape, and discards the rest (see Fig.1). This operation leaves the front end of the signal untouched, but its new peak now lies a distance $d$ ahead of where it used to be. Once the smaller triangle has emerged from behind the screen, the observer sees a smaller signal  advanced relative to the free (no scissors) propagation, and 
moving at the original speed $v_0$. Trying to guess what happened behind the screen, the observer might suspect that the triangle was made to shrink (hence the reduction in magnitude) and also to travel faster, spending there a duration $\tau$ which is $d/v_0$ shorter than the time $L/v_0$ it takes the free signal to traverse the same distance. With $L\approx d$, $\tau$ can be very small, making the 'mean velocity behind the screen', $L/(L/v_0-d/v_0)$, exceed the speed of light. Needless to say,
our example poses no threat to special relativity.
\begin{figure}
\includegraphics[width=7cm, angle=0]{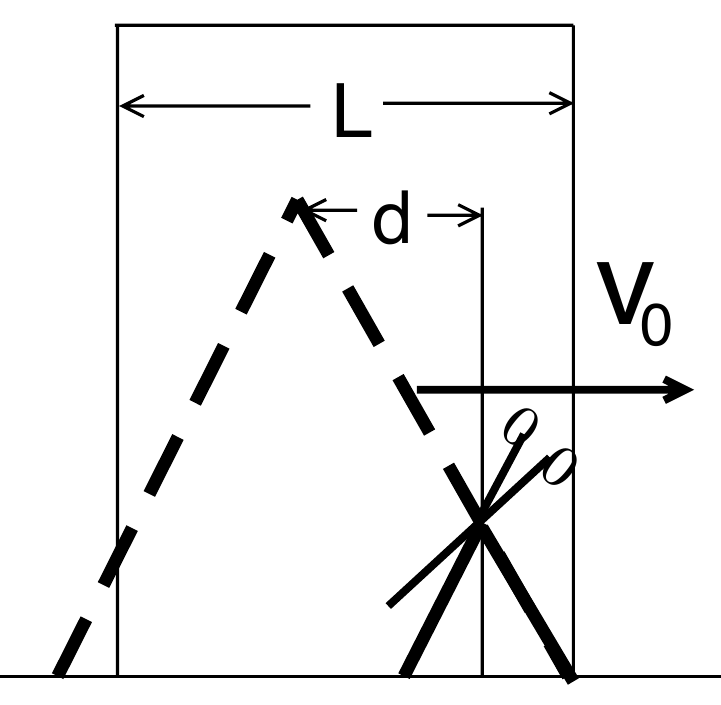}
\label{PROB}
\caption{Primitive reshaping: the front part of the 'signal' moving with a velocity $v_0$ is cut off and the rest (dashed) is discarded. The new peak lies the distance $d$ ahead of the original one and arrives the time $d/v_0$ earlier at a fixed detector. It is wrong, however, to conclude that its mean velocity in the boxed area is $L/(L/v_0-d/v_0)$.
}
\end{figure}
The fallacy is in identifying the  transmitted peak with the incident one, since the causal connection between the two is broken the moment the scissors are brought in to reshape the 'signal'.  
It is also clear that the relevant quantity is the new position of the peak after the truncation, 
from which one can estimate the time of peak's arrival at the detector.
\newline
The analogy with tunnelling of a wave packet is useful, yet incomplete.
Firstly, wave packets are not made of paper, and one would like to know what, if anything, plays the role of the scissors. 
Secondly, it is natural to assume that the 'cut' is made based on the information about the part of the signal which has already passed through the point where the scissors are applied.
Thus, as was pointed out in \cite{WIN2}, if one modifies the part of the signal not yet arrived (dashed line in Fig.1), but leaves the front unchanged, the reshaped signal, unaffected by the modification, will stay the same. And this, depending on the conditions,  may \cite{NATUR} or may not \cite{WIN2}  be observed in an experiment.  We will return to this issue in Sect. XIX.
In the next few Sections we analyse a more subtle reshaping mechanism based on quantum interference.

\section {The model: transmission via entanglement}
Following \cite{DS2} we consider a simple model which, although formally different from tunnelling through a potential barrier, produces a similar effect on the transmitted pulse.
The model
consists of a semiclassical particle of a  mass $\mu$, equipped with a magnetic moment (spin) of $2K+1$ components. The particle  crosses a region $\Omega$ of a width $d$, which contains a weak magnetic field. 
Each spin component encounters in $\Omega$ an additional rectangular potential ($\hbar =1$) $-m \omega_L$, $m=-K, ..0, ...K$,  where $\omega_L$ is the Larmor frequency. 
The mean energy of the particle is much larger then $K\omega_L$ and the incident wave packet is prepared in a product state, which before entering the field propagates with the velocity $p_0/\mu$, $\la x|\Psi(t)\ra=\exp(ip_0x-ip_0^2t/2) G^0(x-p_0t) |a\ra$. Here $G^0(x)$ is the coordinate envelope, whose spreading we will neglect, the spin part has the form $|a\ra \equiv  \sum_{m=-K}^K a_m |m\ra/ \sqrt{N(a)}$, with $a_m$ of our choice, and $N(a)\equiv \la a|a\ra$ the normalisation constant. With some spin components experiencing in $\Omega$ a small potential step, and some a small rectangular well,
the wave packet is split into parts delayed or advanced relative to field-free propagation. By choosing $a_m\equiv 0$ for $m>0$, we can eliminate all advanced components, thus obtaining after crossing the magnetic filed, a final state in which
 \begin{figure}[h]
\includegraphics[width=12cm, angle=0]{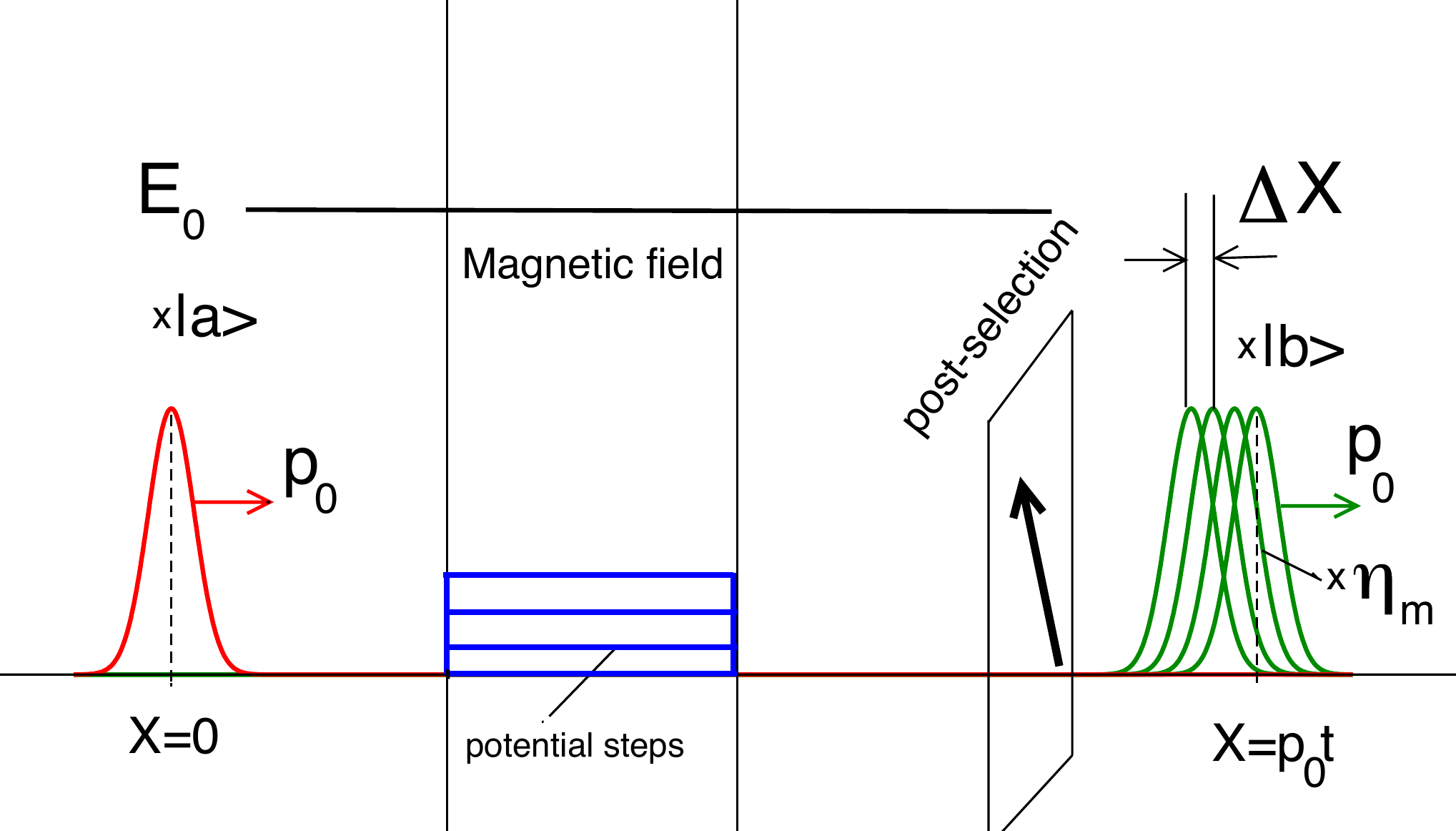}
\caption{(colour online) Schematic diagram of the 'beamsplitter' realised in the model of the Sects. III - IV. Large spin of a fast particle is pre-selected in a state $|a\ra$ prior to entering a magnetic field in which all spin components are delayed. The
spin state is purified upon passing the polariser, after which the coordinate part of the wavefunction is given by a superposition of delayed pulses  weighed by complex quantities $\eta_m$.}
\label{FIG3}
\end{figure}
the translational and spin degrees of freedom are entangled  \cite{DS2} ($\Delta x\equiv \omega_Ld/p_0^2$)
\begin{eqnarray}\label{1}
\la x|\Psi(t)\ra= 
\exp(ip_0x-ip_0^2t/2) \sum_{m=-K}^{0} a_m \times \\
\nonumber
\exp(-im\omega_L d/p_0) G(x-p_0t-m\Delta x) |m\ra/ \sqrt {N(a)}.  
\end{eqnarray}
\section {The model: post-selection and the convolution formula}
We assume further that prior to the particle's arrival at a remote detector we can ensure that its spin is in a state of our choice,  $|b\ra \equiv  \sum_{m=-K}^K b_m |m\ra
/ \sqrt {N(b)}$, e.g., by making it pass through a polariser. Then, on exit from the polariser we have a transmitted pulse whose envelope $G^T$ can, after simple algebra, be written as a convolution \cite{DS2}
\begin{eqnarray}\label{2}
{G^T}(X) = \int_{-\infty}^\infty G^0(X-x')\eta(x') dx'/ \sqrt {N(a)N(b)}\quad\quad \quad
\end{eqnarray}
where $X\equiv x-p_0t/\mu$ and [$\delta(z)$ is the Dirac delta]
\begin{eqnarray}\label{3}
 \eta(x) \equiv \sum_{m=0}^K \eta_m\delta(x+m\Delta x),\quad\\
 \nonumber
  \eta_m\equiv \exp(-im\omega_Ld/p_0)a_mb^*_m .
\end{eqnarray}
Now the transmitted envelope is a superposition of a freely propagating  envelope, plus several of its copies, all delayed, i.e., shifted to the left by $m\Delta x$, $m=0,1,...K$ (see Fig. 2). 
The final location of the particle is determined by 
amplitudes $\eta_m$. These are complex quantities whose real and imaginary parts can be of either sign. Thus, under certain conditions, the constituent envelopes may interfere destructively, 
and produce an 'anomalous' small peak far away from their maxima \cite{AH1}. 
\section {The model: quantum reshaping mechanism}
Free to choose the states $|a\ra$ and $|b\ra$ in Eqs.(\ref{3}), we can design the shape of the transmitted pulse.
Suppose that  we want it to lie a distance $\a>0$ ahead of the freely propagating one,
 and also to have nearly the same  shape as the original wave packet, i.e., ${G^T}(x,t) \approx G^0(X-\a)/ \sqrt {N(a)N(b)}$. 
For this purpose, it would be helpful to have $\eta(x) \sim \delta(x-\a)$, which, clearly,  is only possible, 
provided $\a$ coincides with one of the shifts $-m\Delta x$, $m=0,1,...K$.
We can, however, try
to ensure that the first 
 $K+1$ moments of $\eta(x)$
  are equal to those of $\delta(x-\a)$, 
\begin{eqnarray}\label{4}
\bar{x^n} \equiv \int_{-\infty}^\infty x^n \eta(x) dx=\a^n, \quad n=0,1,...K.
\end{eqnarray}
For $\eta_m$ which may have either sign, 
 Eqs. (\ref{4}) have a non-trivial solution  \cite{DS2} (${\prod'} _{j=0}^K$ indicates   the  product
over all $j\ne m$)
\begin{equation}\label{5}
\eta_m(\a/\Delta x)=
(-1)^{m}\frac{{\prod'} _{j=0}^K(j+\a/\Delta x)}{m!(K-m)!}
\end{equation}
 shown in Fig. 3 for different values of $\a$.
\begin{figure}[h]
\includegraphics[width=14cm, angle=0]{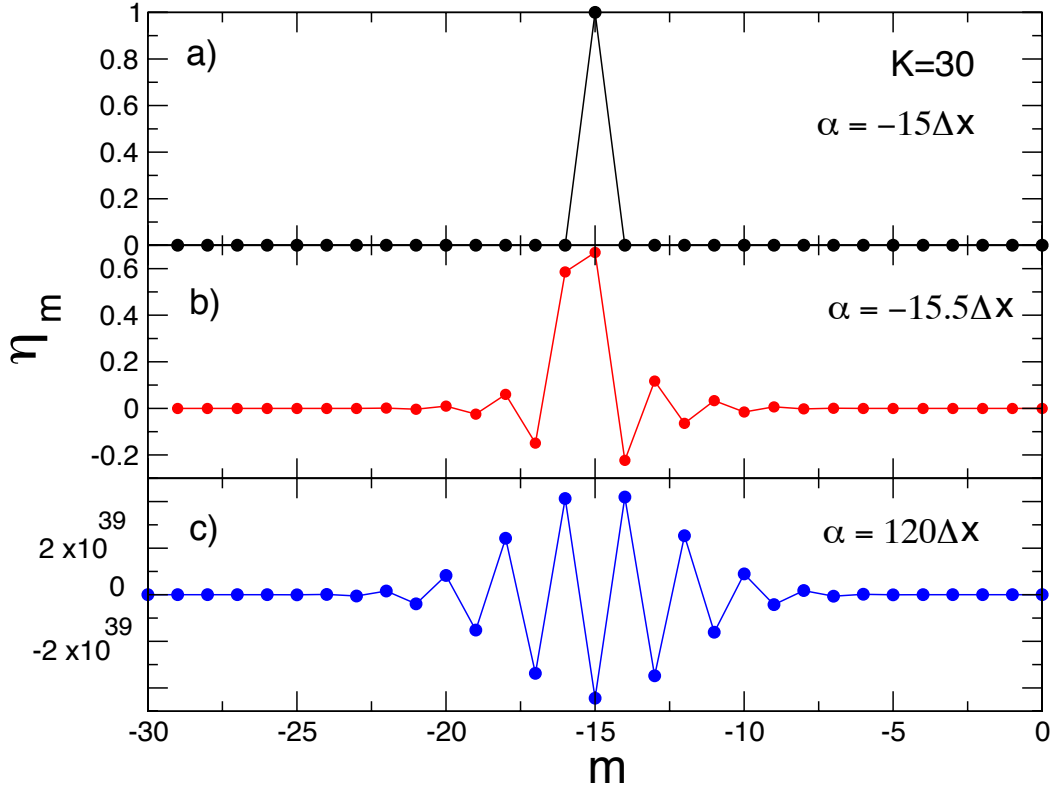}
\label{ETAS}
\caption{ (colour online) The values of $\eta_m$ as given by Eq.(\ref{5}) with $K=30$ \cite{DS2} for: a) $\a$ chosen to coincide with one of the available delays, $\a=-15\Delta x$; b) $\a$ chosen to lie between two of the available delays, $\a=-15.5 \Delta x$;  and c) $\alpha$ chosen so that the transmitted pulse is advanced by $4K\Delta x$ (note the vertical scale).}
\end{figure}
With this choice of the $\eta_m$ we have a curious mathematical object, a collection of $\delta$-functions (\ref{3}), all contained in the region $[-K\Delta x, 0]$, which owing to the approximate  identity
\begin{eqnarray}\label{6}\nonumber
\int G^0(x-x') \eta(x') dx' \approx 
\sum_{n=0}^K \partial^n_xG^{0}(x)/n! \int (x-x')^n \eta(x')dx' \\
= \sum_{n=0}^K \partial^n_xG^{0}(x)(x-\a)^n/n!
\approx G^0(x-\a),\q\q\q\q
\end{eqnarray}
would
act like a single Dirac $\delta (x-\a)$
on any function 
adequately represented by the first $K+1$ terms of its Taylor series.
In a similar way, by choosing $\alpha < 0$ we can make the particle appear to be slowed down. Finally, 
choosing $\alpha$ to be complex valued, $\alpha=\alpha_1+i\alpha_2$, we can effect the translation of  a suitably chosen initial pulse into the complex coordinate plane. Exotic as it may seem, this last case will be needed while discussing the apparent 'superluminality' which arises in tunnelling.

To test the above we may choose the incident pulse to be a Gaussian of a width $\sigma$,
\begin{equation}\label{7}
G^0(x) = (2/\pi \sigma^2)^{1/4} \exp(-x^2/\sigma^2).
\end{equation}
The transmitted pulses for real and complex shifts $\a$ are shown in Figs. 4a and 4b, respectively.
 \begin{figure}[h]
\includegraphics[width=10cm, angle=270]{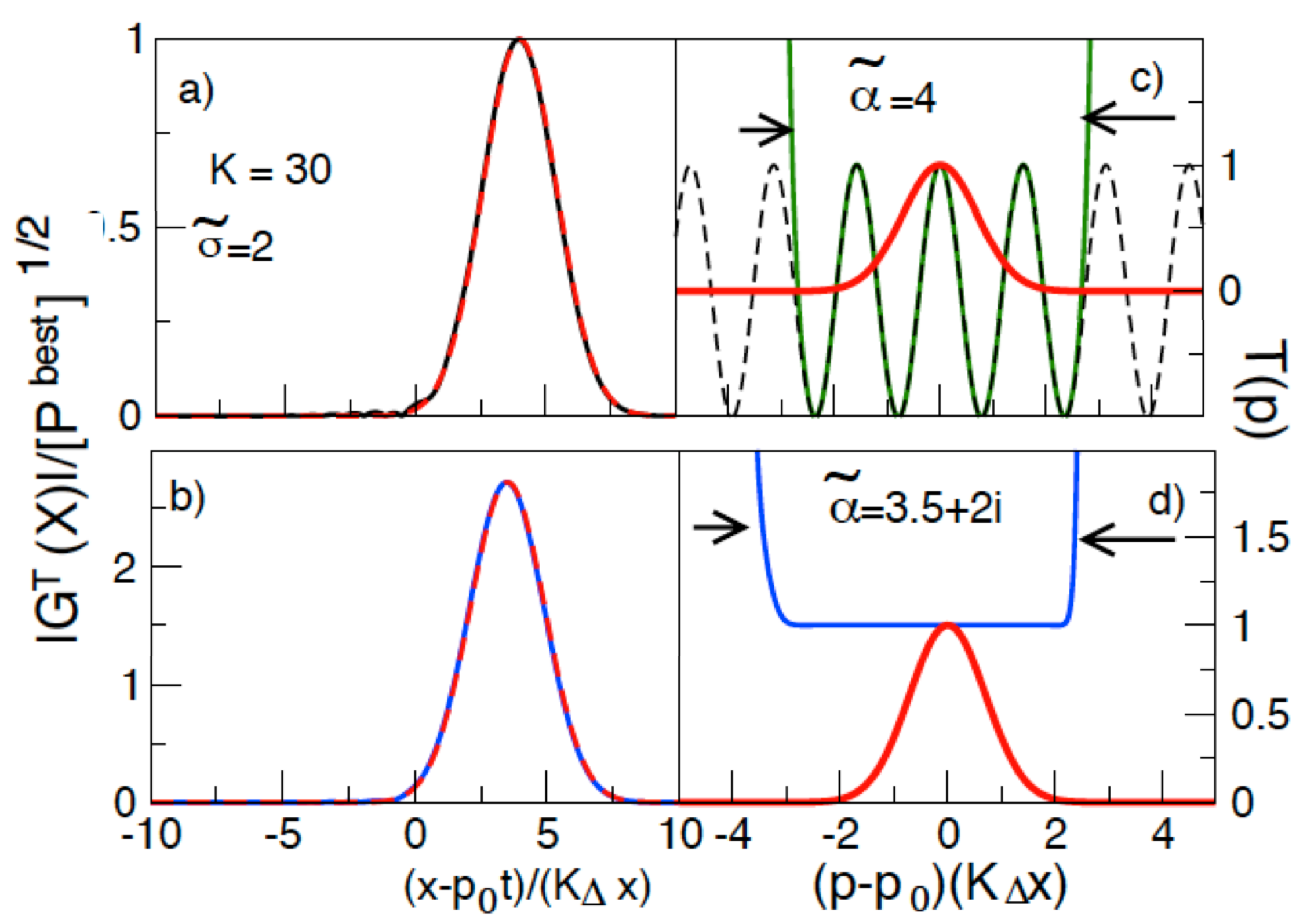}
\label{PACK.pdf}
\caption{(colour online) a) The advanced pulse for the model with $K=30$ and $\tilde{\sigma}\equiv \sigma/(K\Delta x)=2$ and
$\tilde{\a}\equiv \a/(K\Delta x)=4$ (solid) and as given by Eq.(\ref{6}) (dashed); b) Same as (a) but for a complex shift $\tilde{\a}=3.5+2i$. Note that both pulses are amplified by a very large factor $P^{best}$ defined in Eq.(\ref{8}). c) Real part of the transmission amplitude,  $ReT(p)$ for the case (a) (solid). Also shown is $\sin(-\a p)$ (dashed) and the momentum distribution $A(p)$, normalised to unit height (thick solid). The arrows indicate the edges of the superoscillatory band;
d) similar to (c) but for the case (b). The solid line shows modulus of the ratio $T(p)/\exp(-i\a p)$. }
\end{figure}
\newline
In summary, the reshaping mechanism responsible for the advancement of the transmitted wave packet consists in splitting the incident pulse into a number of delayed copies, weighting them with the amplitudes $\eta_m$, and then recombining them to form the transmitted pulse. Note that in our model this last step occurs when the particle passes through the polariser \cite{FOOTW} (see Fig.2).
This has several simple consequences.

\section {The model: causality, speed of information transfer and the success rate}
Our setup acts like a beamsplitter in which the incident pulse, e.g., the Gaussian (\ref{7}), is split into components, none of which overtake the original wave packet. Thus, causality is obeyed explicitly.
\newline
The advanced pulse builds up from the front tails of these components, and
this has well known implications for the speed of information transfer \cite{JAPHA}, 
\cite{DS1}. 
In particular, truncating the incident pulse, say, at the centre and discarding the front part
would eliminate the speed up effect,  as no amplitude will propagate beyond $x=p_0t/\mu$ (for an experimental verification see Ref. \cite{NATUR}). 
Conversely, truncating the pulse so as to leave the front part intact, would not affect the front tails, and the advanced part of the pulse will look as if its rear part had not been amputated.
Accepting, as suggested in \cite{JAPHA}, that the information is transferred by non-analytical features such as cut-offs, one sees that the speed of information transfer never exceeds that of free propagation, $p_0/\mu$.
\newline
Since Gaussian front tails rapidly fall off with the distance, and both $|a\ra$ and $|b\ra$ must be normalised to unity, one expects the advanced pulse to be also reduced in size. 
This means that the probability of successful post-selection $P=1/N(a)N(b)$ is likely to be small,
and only a small fraction of particles would pass through the polariser,
It can be shown \cite{DS2} that the best success rate achieved with the optimal choice 
of the initial and final states $|a\ra$ and $|b\ra$ is 
\begin{equation}\label{8}
\quad   P^{best}(\a) = 1/(\sum_{m=-K}^0 |\eta_m(a)|)^2. 
\end{equation}
For a significant advancement, $\a > K\Delta x$,  the sum $\sum_{m=-K}^0 |\eta_m|$ is a very large number (cf. Fig.3.c)  [unlike $\sum_{m=-K}^0 \eta_m$ which,  according to Eq.(\ref{4}) with $n=0$ always equals unity].
Dependence of the success rate $P^{best}$ on $\a$ is illustrated in Fig. 5 for various real and complex valued spacial shifts. \begin{figure}[h]
\includegraphics[width=12.cm, angle=0]{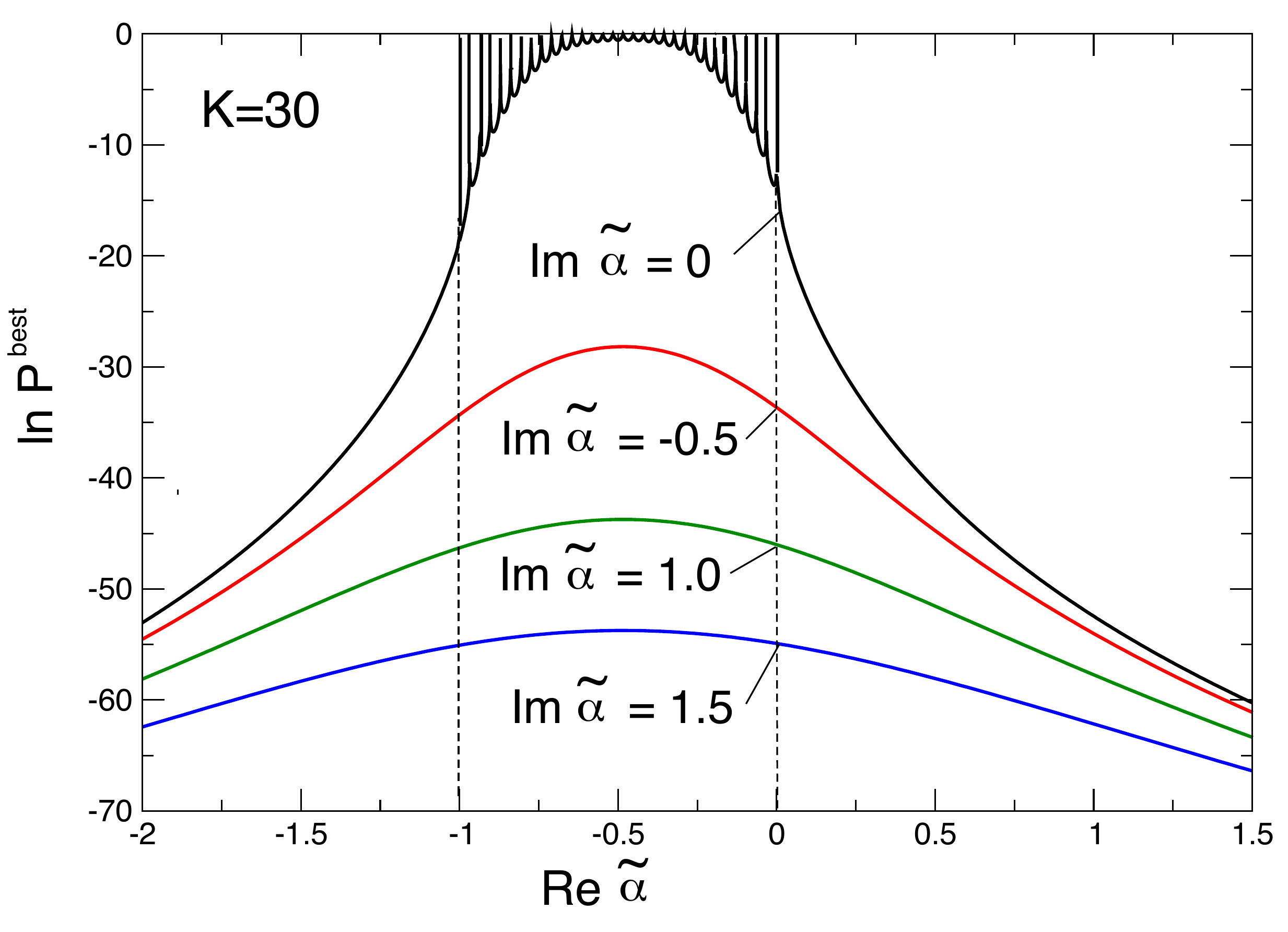}
\label{PROB}
\caption{(colour online) Logarithm of the success probability $P^{best}$ as function of $\tilde{\a}\equiv \a/(K\Delta x)$. The comb-like structure in the case $Im\a=0$ arises because the success rate is unity, whenever $\a$ coincides with one of the available shifts, $-m\Delta X$, $m=0,1,...,K$. In this case Eq.(\ref{5}) just selects the corresponding shape from those shown in Fig. 2.} 
\end{figure}
\section {The model: advancement as a bandwidth phenomenon. Berry's superoscillations}
Alternatively, setting $G^0(x)=\int A(p)\exp(ipx)dp$, we can perform our analysis in the momentum space, by rewriting the convolution  (\ref{2}) as
\begin{eqnarray}\label{11}
{G^T}(X) = \int T(p)A(p)\exp(ipX)dp/ \sqrt {N(a)N(b)},
\end{eqnarray} 
and introducing the transmission amplitude  $T(p)$ as the Fourier transform of  $\eta(x)$,
\begin{equation}\label{12}
T(p) \equiv \int_{-\infty}^{0} \eta(x) \exp(-ipx)dx = \sum_{m=0}^K \eta_m(\alpha) \exp(imp\Delta x).
\end{equation} 
As was shown in Sect.V, for a large $K$,  $\eta(x)$ becomes a good approximation to the $\delta$-function centred at $\alpha$, $\delta(x-\alpha)$. Had this equality been exact, $T(p)$ would have been given by $\exp(-ip\alpha)$. In reality, for $K>>1$, the relation $T(p)\approx \exp(-ip\alpha)$ only holds in a finite region \cite{DS2} ($e$ is the base of natural logarithm)
\begin{eqnarray}\label{14a}
|p|<K/e|\alpha|
\end{eqnarray} 
beyond which $T(p)$ rapidly grows, as is illustrated in Fig. 4c. 

In the case the particle is advanced, $\a>0$,
this behaviour is counterintuitive, since Eq.(\ref{12}) shows that  $T(p)$ builds up from exponentials with only non-negative frequencies $m\Delta x$, $m=0,1,...,K$. 
This is an example of the well known phenomenon of 'supersocillations', where a function locally mimics the behaviour of an exponential with a frequency outside its Fourier spectrum \cite{BERRY1}-\cite{BERRY2}. Thus, in our model, advancement of the transmitted pulse requires that the transmission coefficient have a superoscillatory band (window), and the momentum width of the pulse be small enough for it to fit into the window, as shown in Fig.4c. Whenever this condition is not fulfilled, the transmitted pulse is distorted and the effect disappears. 
The same conclusion applies to the case of a complex shift $\alpha=\alpha_1+i\alpha_2$, shown in Fig. 4d.
\newline
The importance of  superoscillations for the superluminal effect was first suggested in Ref. \cite{SOSC}. The authors of \cite{SOSC} considered a transmission across a comb made of $\delta$-potentials and concluded that the observed advancement of the transmitted pulse does '...result from a superoscillatory superposition at the tail'. We can clarify this statement by noting that an advanced peak results from interference in the front tails, if the problem is studied in the coordinate space. Equivalently, it can be seen to result from superoscillatory behaviour of the transmission amplitude, if the momentum space is used instead.
\section {The model: the 'paradox' and the phase time}
There is, strictly speaking, nothing unusual in the above analysis. Yet, as in Sect. II, we can talk ourselves into a sort of a 'paradox' by the following  (false) reasoning. The pulse in Fig. 4a lies a distance $\alpha$ ahead of the freely propagating one. The only difference between the two pulses is that one had to cross the magnetic field, and the other did not. Thus, the advanced pulse must have spent in the field a duration shorter by 
\begin{equation}\label{9}
\delta \tau =  \mu \alpha /p_0.
\end{equation}
than its free counterpart. 
The free pulse crosses the same region in
\begin{equation}\label{9a}
\tau_0 =  \mu d /p_0,
\end{equation}
hence, the advanced pulse spends there a duration
\begin{equation}\label{9b}
\tau_{phase} = \tau_0-\delta \tau= \frac{\mu }{p_0}(d-\a).
\end{equation}
This parameter is called the phase time since, combining Eqs.(\ref{4}) and (\ref{12}), we can relate it to the momentum derivative of the phase of the transmission amplitude in Eq.(\ref{11}), 
\begin{eqnarray}\label{9c}
\tau_{phase}=\frac{\mu}{p_0}[d+\partial_p\phi (0)],
\end{eqnarray} 
where $T(p)=|T(p)|\exp[i\phi(p)]$.
Finally,  dividing $d$ by $\tau_{phase}$ we should obtain the effective average velocity of the particle in the field, 
 \begin{equation}\label{10}
v_{eff}=p_0/[\mu(1-\alpha/d)].
\end{equation}
Choosing $\alpha=d$ gives $v_{eff}=\infty$, in apparent 'violation' of Einstein's causality. 
Choosing $\alpha=2d$ yields an even more 'paradoxical' result: the duration allegedly spent in the field, $\tau_{phase}=-\mu d/p_0$, is negative.
\newline
Given the simplicity of our model, there is little doubt that the above reasoning is wrong. 
However, to fully dismiss the paradox, one needs  to describe the phenomenon in simple general terms. In the next Section we show that the language for such a description is readily provided by the quantum measurement theory.
\section {The model: resolution of the 'paradox'. Weak measurements}
Returning to Eqs. (\ref{2})-(\ref{3}), we note that they have the same form as those which describe the state of a von Neumann pointer \cite{vN}, employed  to measure an operator with the eigenvalues $-K\Delta x, -(K-1)\Delta x, ..., -\Delta x, 0$ in a $K+1$-dimensional Hilbert space.
The pointer is prepared in a state $G(x)$, and the measured system is pre-selected (prepared) in the state $|a\ra$ and post-selected 
(i.e., found after the measurement)
 in the state $|b\ra$ (see Ref. \cite{AH1} and the Appendix). The analogy is useful.
 \newline
 Firstly, by locating the transmitted particle we conduct a measurement of $x'$, its position (delay or advancement) relative to that of the freely propagating one. The $K+1$ 'eigenvalues' are just the spacial delays experienced by the components of the initial pulse on their pathways across the magnetic field.
\newline
Secondly, the measurement is a quantum one, and its accuracy is determined by the coordinate width of $G^0(x)$, e.g., by the $\sigma$ in Eq.(\ref{7}). An accurate (strong) measurement requires a $\sigma$ as small as possible. But if one chooses $\sigma << \Delta x$, the outgoing pulses in Fig.2 cease to overlap and the speed up effect disappears. Rather, one finds the transmitted particle in one of the positions $p_0t/\mu-m\Delta x$, $m=0,1,...,K$ with the probability $|a_m b_m^*|^2/N(a)N(b)$.
In this case, the $K+1$ pathways lead to different outcomes, with the interference between them completely destroyed.
\newline
To observe the advancement of the transmitted pulse we must, therefore, choose a broader initial Gaussian, one that would satisfy the approximate equalities in Eq. (\ref{6}), and fit into the superoscillatory window in Fig. 4c. This takes us into the limit of highly inaccurate, or 'weak' measurements \cite{AH1}. With $\sigma > K\Delta x$ the pathways remain interfering alternatives. 
According to the Uncertainty Principle \cite{FEYN} such alternatives cannot be told apart, and our attempt to answer the 'Which way?' ('Which delay?) question has no meaningful answer. The answer we do get is the 'weak value' of the spacial delay, $\a$, the first moment (\ref{4}) of the highly oscillatory distribution shown in Fig. 4c. The information about the original $K+1$ sub-luminal delays is hidden from the observer, since the mean of an alternating distribution is not required to lie within its region of support \cite{DSNEGAT},  \cite{DSNEGAT1}. Accordingly, it is our attempt to use the observed shift $\a$ in order to guess the particle's behaviour in the past that leads to the 'paradox' of the previous Section.
\newline 
As was argued by Bohm \cite{BOHM}, an attempt to determine the value of a quantity without destroying interference between the alternatives is fraud with inconsistencies. 
Weak values, observed in experiments designed to leave the interference intact, appear to have no broader meaning, and should not be used to make general assumptions about the observed system. 
It is worth noting that last view is not shared by all authors \cite{AH3}.
 \section {The limited usefulness of the phase time}
 As was discussed in the previous Section, the phase time is an eclectic construction, involving
 quantum weak value of the spacial shift experienced by the particle, $\bar{x}$, and the classical relation between the shift and the duration spent in a potential. 
 Treating it as the duration of a tunnelling event leads to a contradiction.
 Rather, an anomalous weak value serves only to indicate that, with the interference not properly destroyed, a measurement  will give a result we would not normally expect \cite{DSNEGAT}.
Thus, a short or a negative value of the $\tau_{phase}$ suggests that the peaks of the transmitted wave packet may end up ahead of the freely propagating one. This condition is, however, not sufficient. 
To achieve the advancement one needs the superoscillatory  behaviour of the transmission amplitude to persist across certain range of incident momenta, and this depends also on the higher moments of the amplitude distribution $\eta(x)$.
\begin{figure}
\includegraphics[width=14cm, angle=0]{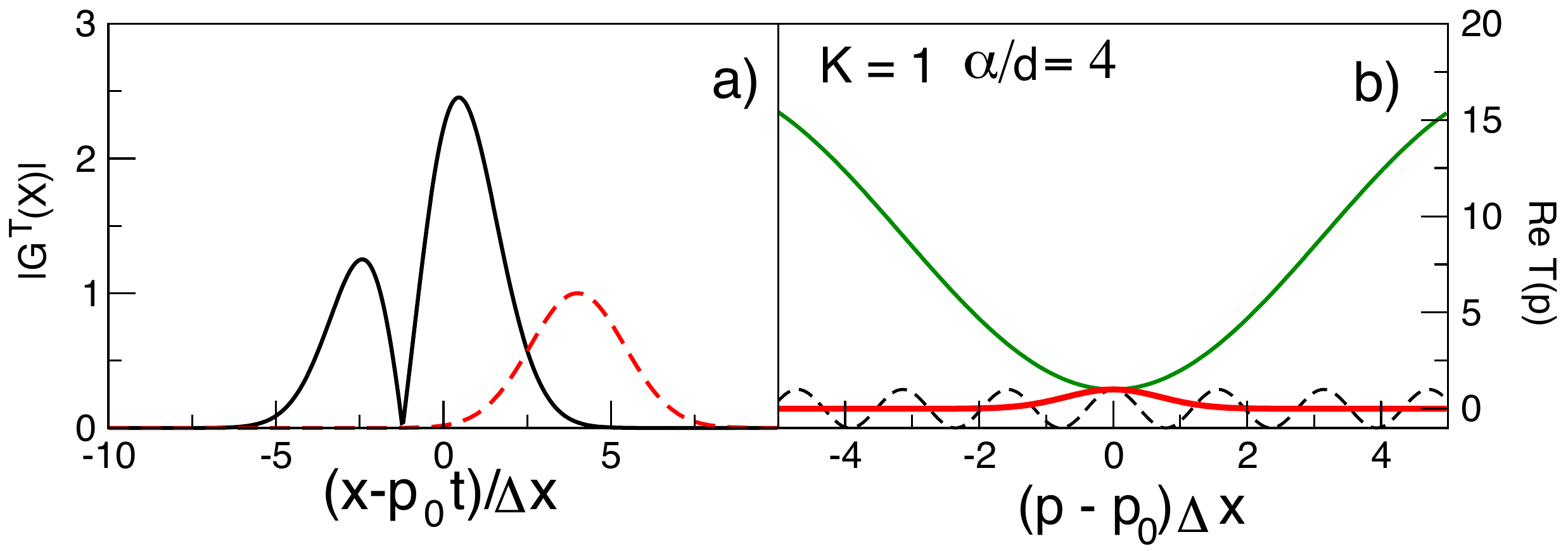}
\label{PROB}
\caption{(colour online) a) Modulus of the transmitted pulse for the model with $K=1$, $\a/d=4$ and $\sigma/d=2$ (solid). Also shown is the pulse shifted by $\a=4d$ (dashed).
b) Modulus of the transmission amplitude, $ReT(p)$, (solid). Also shown are $\sin(-4p)$ (dashed) and  $|A(p)|$ (thick solid).}
\end{figure} 
As an illustration, in Fig. 6 we show the transmission of the initial pulse used in Fig. 4, in a setup tuned so that only the first moment of the $\eta$, $\bar{x}$, equals $4d$. Although $\tau_{phase}$
is the same in both cases, no advancement is achieved in the case shown in Fig. 6, where no well defined supersocillatory band is formed. With this simple observation made, we move on to analyse 'superluminality' in tunnelling.
\section { Tunnelling: the convolution formula and causality}
Our analysis of tunnelling across a classically forbidden region (e.g., a potential barrier or an undersized waveguide) is based on its similarity to the model discussed in Sects. III-X.
 In the momentum space the transmitted pulse is given by
an integral similar to Eq. (\ref{11})
\begin{eqnarray}\label{14}
\Psi^T(x,t) =\int T(p) A(p-p_0)\exp(ipx-ip^2t/2\mu) dp\q
\end{eqnarray} 
where $A(p-p_0)$, peaked at $p=p_0$, is the momentum distribution of the initial pulse,
 and $T(p)$ is the barriers's transmission amplitude.
\newline 
 To confirm that the tunnelling mechanism is causal, we return to the coordinate space.
 There we rewrite Eq. (\ref{14})  in a form similar to the convolution (\ref{2}) by introducing the free propagating pulse, $\Psi^0(x,t) =\int A(p-p_0)\exp(ipx-ip^2t/2\mu) dp$ and two slowly varying 'envelopes', $G^{Z}(x,t,p_0) = exp(-ip_0x+ip_0^2t/2\mu)\Psi^{Z}(x,t)$, $Z=T,0$. 
 As a result, we have
\begin{eqnarray}\label{15}
G^T(x,t,p_0) =T(p_0)\int_{-\infty}^{\infty}G^0(x-x',t,p_0)\eta(x',p_0)dx'. \q\q
\end{eqnarray}
where $\eta(x,p_o)$ is, essentially, the Fourier transform of $T(p)$
\begin{eqnarray}\label{16}
\eta(x,p_0) = [2\pi T(p_0)]^{-1}\exp(-ip_0x)  \int_{-\infty}^{\infty} T(p)\exp(ipx)dp. \q\q
\end{eqnarray}
If the barrier potential does not support bound states, $T(p)$ has no poles in the upper half of the complex momentum plane, and we  have \cite{DS2}
\begin{eqnarray}\label{TS9}
\eta(x,p_0)=\delta(x)+\tilde{\eta}(x,p_0), \quad\q\q 
\end{eqnarray}
where the smooth function $\tilde{\eta}(x,p_0)$ vanishes for positive $x$s (see Fig. 7),
\begin{eqnarray}\label{TS9a}
 \tilde{\eta}(x,p_0)\equiv 0, \quad for \quad x>0. \q\q
\end{eqnarray}
It is easy to check that, just like the $\eta(x)$ in Eq.(\ref{3}), $\int \eta(x,p_0) dx=1$.
\begin{figure}
\includegraphics[width=13cm, angle=0]{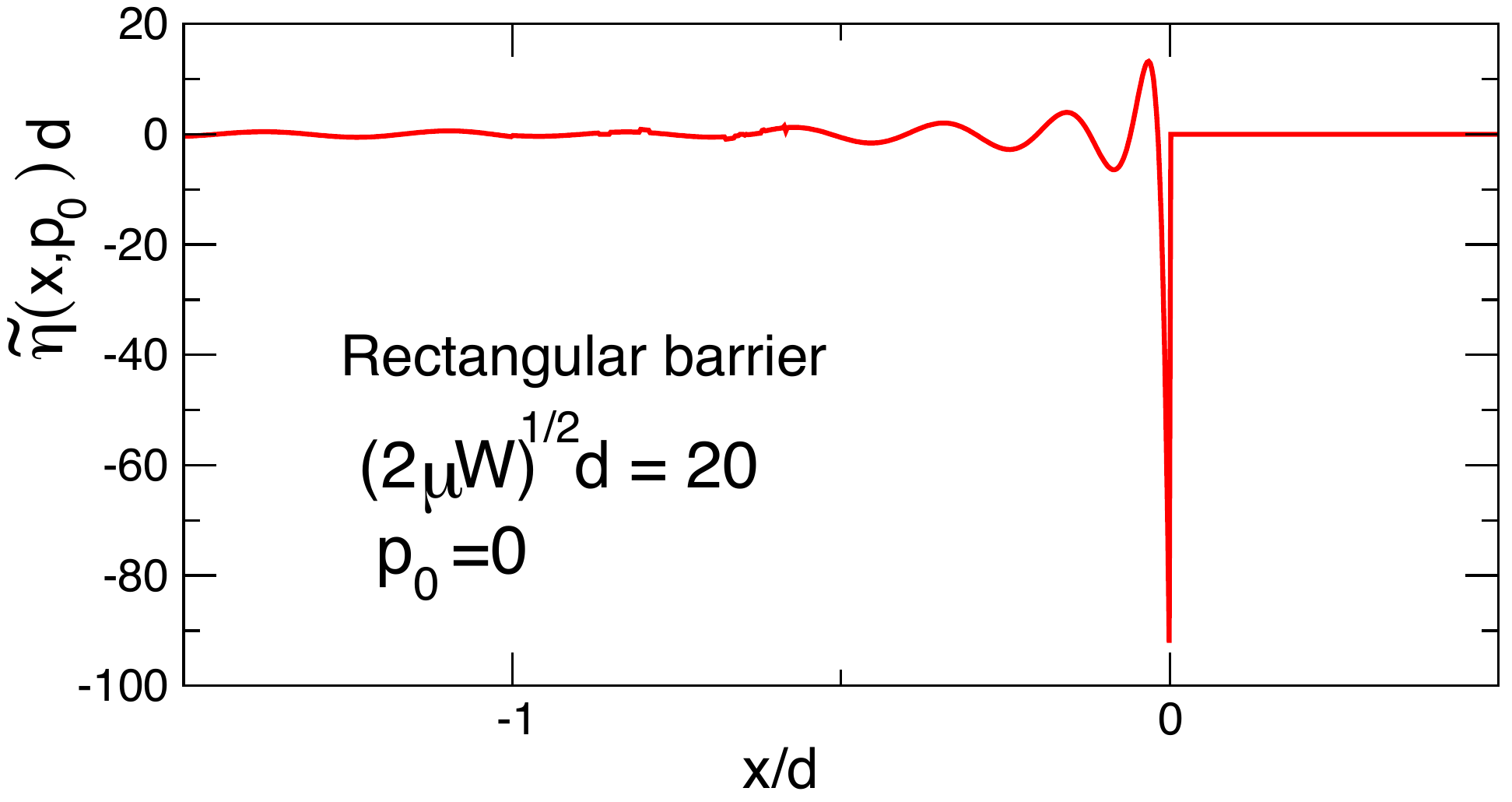}
\label{PROB}
\caption{(colour online) The smooth part of the delay amplitude distribution in Eq.(\ref{TS9}) for a rectangular barrier of a width $d$ calculated numerically from Eq.(\ref{16}) ($Im \tilde{\eta}=0$ since $T(p,W)=T(-p,W)$). Note the sharp negative peak at $x\approx 0$ which serves to cancel the contribution from the $\delta$-function in Eq.(\ref{TS9}) provided the initial pulse is sufficiently broad.}
\end{figure}
The $\delta$-term in Eq(\ref{TS9}) occurs because very fast particles are not affected by the barrier, so that  $T(p)\to 1$ as 
$|p|\to \infty$. 
\newline
Following the analogy with the model of the Sects. II-X, we note that the barrier acts like a beamsplitter with an infinite (continuum) number of 'arms',
 where the transmitted pulse builds from the free envelopes none of which is advanced relative to free propagation. As in Sect. IV the transmission is explicitly causal, since none of the envelopes in  Eq.(\ref{15}) lie ahead of the freely propagating one.
 The position of the peak is determined by the interference between all of  $G^0(x-x',t,p_0)$, and ultimately by the properties of delay amplitude distribution (DAD) $\eta(x,p_0)$ in Eq.(\ref{16}).
 \newline As in Sect. VII, the causality argument can also be made in the momentum space by demonstrating  that the Fourier spectrum of $T(p)$ does not contain negative frequencies. Indeed, 
 we can write $T(p)$ as a Fourier integral
 \begin{eqnarray}\label{24aa}
T(p)=\int_0^{\infty}\exp(ipx)\xi(-x)dx,
\end{eqnarray} 
where $\xi(x)\equiv(2\pi)^{-1}\int_{-\infty}^{\infty}T(p)\exp(ipx)dp$, 
must vanish for $x>0$ 
since $T(p)$, as stated above, has no poles in the upper half of the $p$-plane. Accordingly, an incident plane wave with a momentum $p_0$ 
upon transmission is transformed into a weighted superposition of plane waves with all possible backward shifts, 
 \begin{eqnarray}\label{24ab}
\exp(ip_0 x) \to T(p_0)\exp(ip_0x)=\\ 	\nonumber\quad\q\q 
\int_{-\infty}^0 \xi(x')\exp[ip_0(x-x')]dx'.
\end{eqnarray} 
With Eq.(\ref{15}) in place, we can consider wave packet propagation from the point of view of quantum measurement theory.
\section {Tunnelling: self-measurement of the delay}
We start by noting first several differences between tunnelling and the entanglement-based transmission of the Sect.III. Tunnelling does not involve any external degree of freedom, such as the spin variable used in Sect. III. We do not have the flexibility in choosing the DAD (\ref{TS9}) as we wish, since its properties are now determined by the barrier potential. 
It tunnelling, there is post-selection, albeit in a slightly different sense. In Eq.(\ref{14}) we have already post-selected  the  particles which are transmitted, and discarded those reflected by the barrier. 
\newline
The rest of analysis is remarkably similar. We have an amplitude distribution $\eta(x',p_0)$ of spatial delays ($x'$) for  a particle whose momentum is precisely $p_0$ [cf. Eq. (\ref{24ab})].  For such a particle all delays interfere
and, finding it at a location $x$, one learns nothing about which delay, $x'$, it has actually experienced, except that causality ensures that $x'\le 0$ for all the pathways involved. 
\newline
For a particle whose initial envelope $G^0(x)$ is sharply peaked around $x=0$ with a width $\delta x$, the situation is different. If the spreading of the free wave packet can be neglected, 
finding the particle in $x$ suggests that the shift almost certainly lies in the region $[x-p_0t/\mu-\delta x,x- p_0t/\mu +\delta x]$. Thus, by localising the transmitted particle,  we perform a quantum measurement of the spacial delay at $p_0$, to the accuracy $\delta x$. Note that in the wave packet scattering  the particle itself plays the role of a von Neumann pointer.
As in Sect. IX, the measurement is accurate, or 'strong', for a wave packet narrow in the coordinate (broad in the momentum) space, and inaccurate, or 'weak', 
 if it is broad in the coordinate (narrow in the momentum) space. 
 \newline
Finally, the presence of the singular term, $\delta(x)$, in Eq. (\ref{TS9}) indicates that in an ideal highly accurate measurement, $\delta x \to 0$, one would always find a zero delay, $x'=0$. Indeed, a wave packet
narrow in the coordinate space contains high momenta, most of which are not affected by the presence of the barrier, and we correctly recover the free motion result. Note that a similar behaviour of an amplitude distribution accounts for Zeno effect in quantum measurement theory \cite{DS5}.
\section {Tunnelling: 'superluminal' advancement. Superoscillatory behaviour}
Like most authors \cite{REV}-\cite{REV3}, we consider transmission of a Gaussian wave packet 
of a coordinate width $\sigma$, located at $t=0$ a distance $x_0$ to the left of a rectangular barrier, 
\begin{eqnarray}\label{17}
V(x)=W\q for \q 0\le x \le d \q and \q 0 \q otherwise.\q\q
\end{eqnarray}
The freely propagating envelope of such a pulse, required in Eq.(\ref{15}), has the form (see, e.g., \cite{DS3})
\begin{eqnarray}\label{17a}
G^0(x,t,p_0)=[2\sigma^2/\pi\sigma_t^4]^{1/4}\exp[-(x-p_0t+ x_0)^2/\sigma_t^2],\q\q
\end{eqnarray}
which differs from Eq.(\ref{7}) in that now we include the spreading of the wave packet, whose complex valued 'width',  $\sigma_t^2\equiv \sigma^2+2it/\mu$ depends on time. 
It is convenient to write the  transmission coefficient $T(p)$  as a geometric progression \cite{DS3}
\begin{eqnarray}\label{18}
T(p,W) = \quad \quad \quad \quad \quad \quad \quad  \quad \quad \quad \quad \quad \quad\\
\nonumber
 \frac{4pk \exp[-i(p-k)d]}{(p+k)^2}\sum_{n=0}^{\infty}\frac{(p-k)^{2n}}{(p+k)^{2n}}\exp(-i2nkd),
\end{eqnarray}
where $k\equiv(p^2-2\mu W)^{1/2}$.  Then only the first term needs to be retained if the barrier is sufficiently broad. We note that the factor $\exp[-ipd]$ in Eq.(\ref{18}) has a 'superoscillatory' aspect since, as was shown in Sect. XI, for a barrier the Fourier transform of $T(p,W)$ contains only exponentials $\exp(ipx)$ with positive frequencies, $x>0$. It is easy to guess that it is this
factor which is responsible for the advancement of the tunnelled pulse by the barrier width $d$, as will be shown below.
\newline
Well above a broad barrier Eq.(\ref{15}) yields the classical result \cite{DSCLASS}.
For $p_0^2/2\mu>>W$, $(p_0-2/\sigma)^2/2>>W$, the highly oscillatory $\eta(x,p_0)$ in Eq.(\ref{15}) develops a stationary region near $x'=x'_s\equiv \mu d(p_0^{-1}-k_0^{-1})< 0 $, which selects a single delayed shape $G^0(x-x_s,t,p_0)$ from the collection of retarded envelopes in Eq.(\ref{15}).
This recovers the classical result: a particle passes over a potential hill with a reduced velocity and, therefore,  lags behind the free one.
\newline Well below the barrier, $p_0^2/2<W$, $(p_0+2/\sigma)^2/2<W$, $(p_0-2/\sigma)^2/2>0$
a simulation shown in Fig. 8 finds that the tunnelling pulse is reduced by the factor $|T(p_0)|$, and the peak of the transmitted probability lies approximately the barrier width $d$ ahead of the freely propagating one. 
\begin{figure}
\includegraphics[width=14cm, angle=0]{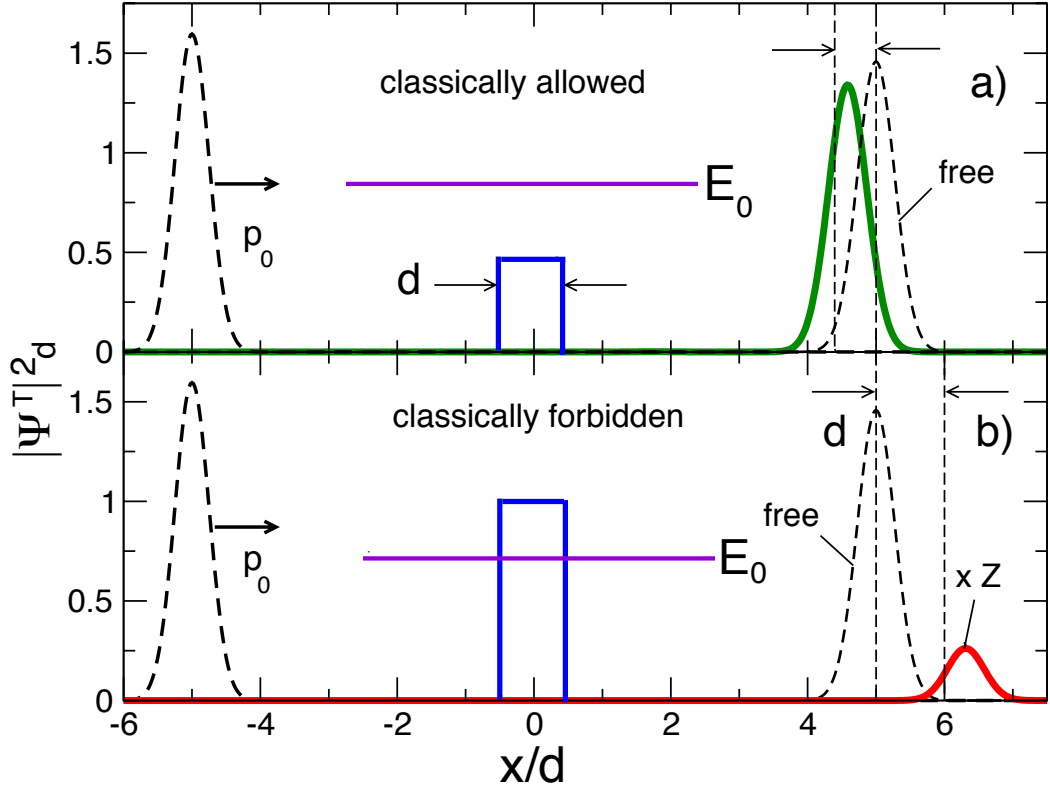}
\label{PROB}
\caption{(colour online) The same initial wave packet (left) propagates, for the same time:
a) passing above a rectangular barrier. In the barrier region the peak moves slower, and the particle is delayed; b) By tunnelling. The transmitted pulse is greatly reduced and its peak lies ahead of the free one. The solid curves are obtained by numerical evaluation of the integral (\ref{14}), the dashed lines correspond to free propagation, and
$Z=\exp[(2\mu W-p_0)^{1/2}d]$ is a large parameter.}
\end{figure}
As in Section IX, we do not take this as a proof that the particle has traversed the barrier infinitely fast. Rather, we concentrate on the weak 'measurement' responsible for this counterintuitive result.
\section {Tunnelling: complex spacial 'delays' }
As in Sect. V, the transmitted pulse builds up from the front tails of the retarded envelopes in the region where  the DAD $\eta$ has no support. This suggests inspecting the moments of $\eta(x,p_0)$, in order to see whether their behaviour is similar to that seen in Eq.(\ref{4}). We have 
 \begin{eqnarray}\label{19}
\bar{x^n} \equiv \int_{-\infty}^{\infty} x^n\eta(x,p_0)dx=i^n\partial^n_p T(p)/T(p)|_{p=p_0},
\end{eqnarray}
where the second equality follows from the Fourier relation (\ref{16}). 
For a broad barrier, $p_0d>>1$ (tunnelling across a high barrier, $W/p_0^2>>1$ can be treated in a similar way  \cite{DS3}) we find
 \begin{eqnarray}\label{20}
\bar{x^n} = \alpha^n+O(d^{n-1}),\q \alpha \equiv d\left (1+\frac{ip_0}{\sqrt{2W-p_0^2}}\right).\q
\end{eqnarray}
Increasing the width of the barrier, $d\to\infty$, and ignoring for the moment in Eqs. (\ref{20}) the corrections of order $O(d^{n-1})$, we conclude that
tunnelling tends to shift the original pulse 
 into the complex coordinate plane  by $\alpha$, and at the  same time reduce its magnitude by the factor of $T(p_0)$ [cf. Eq.(\ref{15})], 
\begin{eqnarray}\label{21}
G^T(x,t,p_0) \approx T(p_0)G^0(x-\alpha,t,p_0).
\end{eqnarray}
From this we can obtain the 
 time at which the peak of the transmitted probability density, $P^T(x,t,p_0)\equiv |G^T(x,t,p_0)|^2$,  arrives at a given location.
Neglecting also the effects of spreading, i.e., assuming $\sigma_t\approx \sigma$,   for the Gaussian pulse (\ref{17a})
 we have 
\begin{eqnarray}\label{21a}\nonumber
P^T(x,t,p_0)\approx[2/\pi \sigma^2]^{1/2}|T(p_0)|^2\exp(2Im\a^2/\sigma^2)\times \\ 
\exp[-2(x-p_0t+x_0-Re \a)^2/\sigma^2].\q\q
\end{eqnarray}
Thus, the peak of the tunnelled probability density lies a distance $\approx Re\a$ ahead of the freely propagating one, 
 and would arrive at a fixed detector a time $\delta\tau$ earlier
 \begin{eqnarray}\label{22}
\delta \tau \approx \mu Re \a/p_0= \mu d/p_0
\end{eqnarray}
 than it would do by free propagation.
\newline
Finally, taking into account the spreading, i.e., the time dependence of $\sigma_t$ in Eq. (\ref{17a})
and recalculating $P^T$ in Eq. (\ref{21a}) adds another speed up effect, also contained in a compact form in Eq.(\ref{21}). The effect, known as 'momentum filtering' \cite{REV}, consists of  increase of the initial mean momentum $p_0$ by the amount
 \begin{eqnarray}\label{23a}
\Delta p_0 =2 Im \a/\sigma^2,
\end{eqnarray}
which gives the transmitted pulse also a boost in velocity responsible for the additional advancement of the pulse in Fig. 8b (for details see Ref. \cite{DS3}).
We note also that a relation similar to Eq. (\ref{23a}) exists in the quantum measurement theory between the imaginary part of a weak value and the change in the momentum of the von Neumann pointer (see Ref. \cite{AH3}, p.412).
\section {Tunnelling:  the Hartman effect and 'sharp' weak measurements}
Equation (\ref{22}) of the previous Section may suggest that by making a barrier broader, one would be able 
to advance the transmitted pulse by an ever larger distance $\sim d$. 
This so-called Hartman effect is usually formulated by constructing from Eq. (\ref{22}) a phase time similar to (\ref{9c}), 
 \begin{eqnarray}\label{23}
 \tau_{phase}= \frac{\mu}{p_0}[d+\partial_p \phi(p_0)],
\end{eqnarray}
where, as before, $T(p)=|T(p)|\exp[i\phi(p)]$. 
It is then argued that this time (which, as we have shown above, is not really a time) does not increase with the barrier width, $\tau_{phase}\sim O(1)$ for $d\to \infty$. 
As discussed in Sect. X, this alone does not guarantee apparently 'superluminal' advancement.
Indeed,  Hartman \cite{HART} has found that a given wave packet does exhibit 'superluminal' advancement for certain barrier widths $d$, but for broader barriers the effect disappears, 
as tunnelling becomes negligible, and the transmission becomes dominated by the momenta which pass over the barrier top. The discussion about whether the Harman effect and its variants do exist in the limit $d\to \infty$ continues in the literature to date \cite{HEFFECT1}-\cite{HEFFECT4}.
\newline
The tunnelling regime can be preserved by increasing, as the barrier become broader, also the coordinate width of the wave packet, $\sigma$. Indeed, in the plane wave limit, $\sigma \to \infty$, the pulse always probes the superoscillatory transmission amplitude, since from Eq.(\ref{18}) we  have [cf. also Eq.(\ref{24ab})]
 \begin{eqnarray}\label{23aa}
\exp(ip_0x)\to \sim |T(p_0)|\exp[ip_0(x-d)].
\end{eqnarray}
One question is whether the width $\sigma$ must remain larger than the advancement $d$, in which case one has to collect statistics of many delayed and advanced arrivals in order to observe the forward shift of the transmitted peak. If, on the other hand, it is possible to also ensure $\sigma << d$, then already the first early arrival at the detector would confirm the 'superluminal' effect. Since the tunnelling probability of a broad barrier is small, the second case is clearly more favourable. 
\newline
A similar situation occurs in the 'weak' quantum measurements \cite{AH1}-\cite{AH3}, if one's aim is to measure an 'anomalous' weak value (see Appendix). 
There an anomalous mean value outside the spectrum of the measured operator $\ah$  occurs due to interference in the tails of the Gaussians centred at the eigenvalues of $\ah$. This requires broad Gaussians, which leads to the problem just discussed. 
In some cases it is possible to construct  a measurement where already the first successful trial would confirm the 'anomalous' result with certainty. The authors of Refs. \cite{AH2} and \cite{AH3} called such sharp weak measurements 'not really weak' and gave some recipes for their preparation. A more detailed discussion of the  analogy between weak measurements and wave packet transmission can be found in Ref. \cite{DS4}.
\newline
In the next Section we show that for $d\to \infty$ the Hartman effect exists in the sense of such  'sharp' weak measurement, allowing a patient observer confirm the effect already with the first 
arriving particle.
\section {Tunnelling: Hartman effect for infinitely wide barriers}
We recall first that our derivation of Eq. (\ref{21}) in Sect. XIV is incomplete, since in  Eqs. (\ref{20}) we have omitted corrections to $\bar{x^n}$ which are of order of $d^{n-1}$ and, strictly speaking, not negligible.
Because of these corrections, the barrier transmission amplitude $T(p)$ does not have a well defined super-oscillatory window similar to the one shown in Fig. 4d. 
\newline
The proof of the last statement of the previous Section, given in \cite{DS4}, is based on a simple estimate. 
An exponential $\exp[d\sum_{n=0}^{\infty}f^{(n)}\sigma^n]$, 
 with $d\to\infty$, $\sigma\to 0$ and $f^{(n)}\sim 1$ can be approximated by $\exp(f^{(0)}d)\exp(\sigma f^{(1)}d)$ provided $d>>1$, $d\sigma\sim 1$ and $\sigma^nd<<1$ for $n>1$.
 \newline
For a broad barrier the transmission amplitude (\ref{18}) is proportional to $\exp[-id(p-k)]$,
and we may expand the exponent, $f\equiv p-ik$, in a Taylor series around $p=p_0$. For a Gaussian wave packet $A(p)$ in Eq.(\ref{14}) is proportional to $\exp(-p^2\sigma^2/4)$. Expecting the typical value of $(p-p_0)$ to be $\sim 1/\sigma$, we estimate the terms in the exponent of $T(p)$ as
 \begin{eqnarray}\label{25}
d\sum_{n=0}f^{n}(p_0)(p-p_0)^n \sim f(p_0)d+\frac{d}{\sigma} f'(p_0)+\\ \nonumber
\sum_{n=2}\frac{d}{\sigma^n} f^{n}(p_0)/n!.
\end{eqnarray} 
Thus, provided
 \begin{eqnarray}\label{26}
\sigma = const\times d^{\frac{1+\epsilon}{2}}, \q 0 < \epsilon\le1 ,
\end{eqnarray} 
the last sum in Eq.(\ref{25}) can be neglected, and the incident pulse would 'see', as $d\to \infty$, the approximate transmission amplitude 
 \begin{eqnarray}\label{27}
T^{app}(p)=
  T(p_0)\exp[-i\a(p-p_0)].\q
\end{eqnarray} 
Inserting Eq. (\ref{27}) into Eq. (\ref{14}), for such wave packets we recover Eqs. (\ref{21}) and (\ref{21a}), so that the peak of the transmitted density lies approximately the distance $d$ ahead of a 
freely propagating one, with the spread of the tunnelled particle's position, $\sigma$ being much smaller, $\sigma/d \approx const/d^{\frac{1-\epsilon}{2}}$ than the 'superluminal' shift $d$. This is illustrated in Fig. 9 for various barrier widths.
\begin{figure}
\includegraphics[width=14.cm, angle=0]{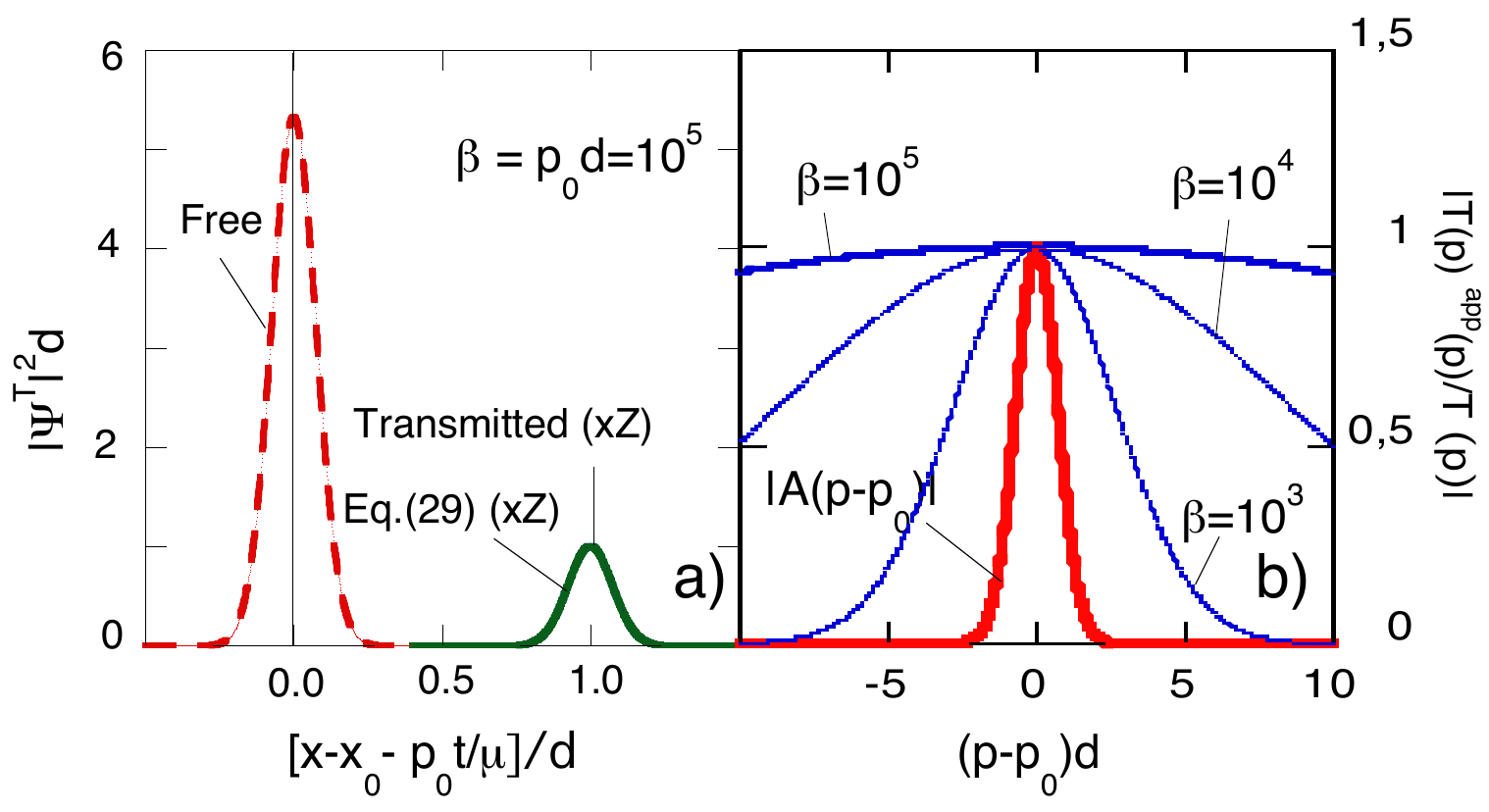}
\label{PROB}
\caption{(colour online) a) Tunnelling of a Gaussian wave packet  across a broad rectangular barrier, $p_0d=10^5$, $p_0^2/2\mu W=0.25$, $\epsilon = 1$, $\sigma/d=0.15$, $p_0t=1.5d$, and $x_0=-3\sigma$.
Transmitted probability is multiplied by a large factor $Z=\exp(k_0d)/(4\times10^{12})$.
b) The ratio between $T^{app}(p)$ in Eq.(\ref{27}) and the exact  transmission amplitude $T(p)$ for different values of $\beta=p_0d$, 
while $p_0^2/2\mu W$ and $\sigma/d$ are kept constant. As the barrier becomes wider, the ratio tends to unity for all initial momenta, 
 and the transmitted envelope becomes a reduced copy of the original one shifted by $\a$ in Eq.(\ref{20}) into the complex $x$-plane. The momentum distribution $A(p)$ (thick solid) is normalised to unit height.
}
\end{figure}
Further numerical examples can be found in Ref. \cite{DS4}, together with a similar recipe for constructing
weak von Neumann measurements designed to give, with near certainty, an unusual weak value
of the measured quantity.
\section {Temporal delays}
Ultimately, one is interested in how soon the transmitted particles will arrive at a fixed detector or, more generally, in the time variation if the transmitted amplitude at  the point of detection, $x_{det}$.
Once the shape of the pulse is known, this information is easily recovered by treating in Eq.(\ref{15}) $x=x_{det}$ as a parameter and $t$ as a variable. This shows that 
 \begin{eqnarray}\label{28}
G^T(x_{det},t) =T(p_0)[G^0(x_{det},t,p_0)+\q\q\q\\ \nonumber 
\int_{-\infty}^{0}G^0(x_{det}-x',t,p_0)\tilde{\eta}(x',p_0)dx']
\end{eqnarray} 
 builds up as the sequence of spacially retarded pulses passes through $x_{det}$. 
\newline
Alternatively, we can follow Ref. \cite{JAPHA}, in trying to obtain a decomposition of the amplitude at $x_{det}$ into retarded components by changing in Eq.(\ref{14}) from the momentum to the energy representation, and writing the result as a different convolution, 
\begin{eqnarray}\label{29}
G^T(x_{det},t) =T(E_0)\int_{-\infty}^{\infty}G^0(x_{det},t-t',E_0)\zeta(t',E_0)dt', \q\q
\end{eqnarray}
where 
\begin{eqnarray}\label{30}
\zeta(x,E_0) = [2\pi T(E_0)]^{-1}\exp(iE_0t) \\ \nonumber
 \int_{-\infty}^{\infty} T(E)\exp(-iEt)dE. \q\q
\end{eqnarray}
Equations (\ref{29})-(\ref{30})
are identical to 
Eqs. (\ref{28}) and (\ref{16}) for a dispersion-less medium, $E(p)=cp$,  where a wave packet retains its original shape, and the signal from a pulse delayed by $x'$ in space, is just the one delayed in time by $x'/c$. For a non-relativistic quantum particle  or a photon in a narrow waveguide one has $E(p)=p^2/2\mu$, and the
 transmission amplitude $T(E)$ may have poles on both sheets of its two-sheet Riemann surface.
This makes both  evaluation of $\zeta(t,E_0)$ and presenting the causality argument less straightforward.
A detailed analysis will be given elsewhere. For the present purpose it suffices to note that the analysis in terms of the space shifts, adopted in this paper, appears to be a simpler tool for investigating the 'superluminal' paradox.
\section {Relation to other quantum times}
In Sections VIII and X we have argued that apparent 'superluminality' has to do with spatial reshaping of the incident pulse, and should not be used to infer the duration a tunnelling particle spends in the barrier. There is, however, a time variable, called the traversal (or Larmor) time \cite{BAZ}-\cite{BAZBOOK}, which is constructed to yield precisely this duration. Next we briefly compare these two variables. 
\newline
The traversal time can be defined as a functional on virtual Feynman paths, which yields the duration $\tau$ a path spends in a given region of space $\Omega$ \cite{TT2}. Summing Feynman amplitudes $\exp(iS)$ \cite{FEYN} over the paths with a given value of $\tau$ yields the transition amplitude, $\varphi(\tau)$, 
with additional condition that the particle spends
$\tau$ seconds in the $\Omega$,
e.g., in the barrier region.
It can be shown (see, for example, \cite{TT2}) that for a non-relativistic particle with a momentum 
$p_0$, incident on a rectangular barrier of a height $W$ this amplitude is given by
\begin{eqnarray}\label{Y1}
\varphi(p_0,\tau)= (2\pi)^{-1}\int_{-\infty}^{\infty} dV T(p_0,W+V) \exp(iV\tau).\q
\end{eqnarray}
The Fourier transform (\ref{Y1}) suggests an uncertainty relation (we restore $\hbar$ temporarily)
\begin{eqnarray}\label{Y2}
\Delta \tau \Delta V \gtrsim \hbar/2
\end{eqnarray}
which implies that in order to know the duration spent in the barrier,  we must somehow introduce an uncertainty in the barrier height. This is precisely what is achieved by coupling the particle to a Larmor clock \cite{BAZ}-\cite{TT2}, where each spin component encounters a different barrier or well. For large spins and not-too-small fields $\Delta V$ can be made large. Then $\tau$ is determined accurately, yet tunnelling is seriously perturbed, or even destroyed by the measurement \cite{TT2}. With small spins and vanishing fields one avoids the perturbation, but evaluates the weak value of the traversal time functional. Such are the Larmor times obtained in Refs. \cite{BUTT}, \cite{BASK}.

As far as we know, there is no functional representing the spacial delay in Eq. (\ref{15}), yet the rest of the analysis can be conducted in a similar manner. From Eq. (\ref{16}) we have
\begin{eqnarray}\label{Y3}
\eta(x,p_0) \sim (2\pi)^{-1}) \int_{-\infty}^{\infty} T(p+p_0)\exp(ipx)dp, \q\q
\end{eqnarray}
and a position-momentum uncertainty relation 
\begin{eqnarray}\label{Y4}
\Delta x \Delta p_0 \gtrsim \hbar/2.
\end{eqnarray}
Equation (\ref{Y4}) demonstrates that to know the spacial delay or advancement of the 
transmitted pulse we must introduce an uncertainty in the pulse's momentum.
This is achieved by sending in a peaked wave packet, rather than a plane wave with no obvious reference point.
With a pulse narrow in space, the delay is determined accurately, but tunnelling is 
seriously perturbed as most of the momenta now pass above the barrier.
With a broad pulse one avoids the perturbation, but evaluates the weak value of the shift and from it derives the phase time (\ref{23}). 

We have, therefore, two quantities, one is a duration 'conjugate' to the barrier height, 
the other is a distance, 'conjugate' to the particle's momentum. 
 One is relevant for problems where 
a static barrier is modified, e.g., by a small external field, the other relates to wave packet propagation.
There has been some confusion as to which of them represents the 'true' tunnelling time, since both approaches give the same result in the (semi)classical limit but differ in the full quantum case. For example, Baz' \cite{BAZBOOK}, having evaluated the weak Larmor time, criticised the phase time obtained by Smith \cite{SMITH} for being incorrect. In fact, the two are complimentary quantities which should not be in competition with each other. 
\section {Is there a simpler physical explanation?}
One outstanding question concerns the physical origin of the superluminal effect.
Several authors have made efforts toward answering it.
\newline 
Nimtz and co-workers suggested that 'superluminality' in propagation of electromagnetic pulses could be explained in terms of virtual photons capable of violating Einstein relativity on a microscopic scale \cite{NIM3}- \cite{NIM5}. Their explanation appears to rely on the fact that relativistic one-particle propagators decay, yet do not vanish identically, outside the light cone. 
\newline 
Winful, \cite{REV3}, \cite{WIN1}-\cite{WIN3} rejected the suggestion that causality could be violated, since evanescent waves are described by classical Lorentz-invariant Maxwell equations. Instead, he argued, the effect could be explained in terms of the energy (or probability) stored with exponentially decaying density in the classically forbidden region, where no actual propagation of the pulse occurs. Rather, the energy spills out at the right end of the barrier shortly after being pushed by more incoming energy at its left end.
\newline Buettiker and Washburn  \cite{BUTT1} also dismissed speculation about superluminal velocities by pointing out, following Refs.  \cite{RESH1} and \cite{JAPHA}, that the transmitted pulse is shaped out of the front end of the incident one, in a manner similar to what is shown in Fig.1.
Their argument was disputed by Winful \cite{WIN2}, on the ground that by carving the front part of a two-humped pulse one should get a single-humped transmitted pulse, whereas in an experiment the transmitted signal repeats the original two-humped shape. 
\newline
It is easy to show that the objection in Ref. \cite{WIN2} is not valid, since the reshaping mechanism relies on the superposition principle \cite{DS4a}. 
Indeed, one can construct an initial envelope 
$G(x)$ with  two (or many) humps by adding  $J$ Gaussians of the type (\ref{7}), all shifted in space, (see Fig. 10)
\begin{figure}
\includegraphics[width=10cm, angle=0]{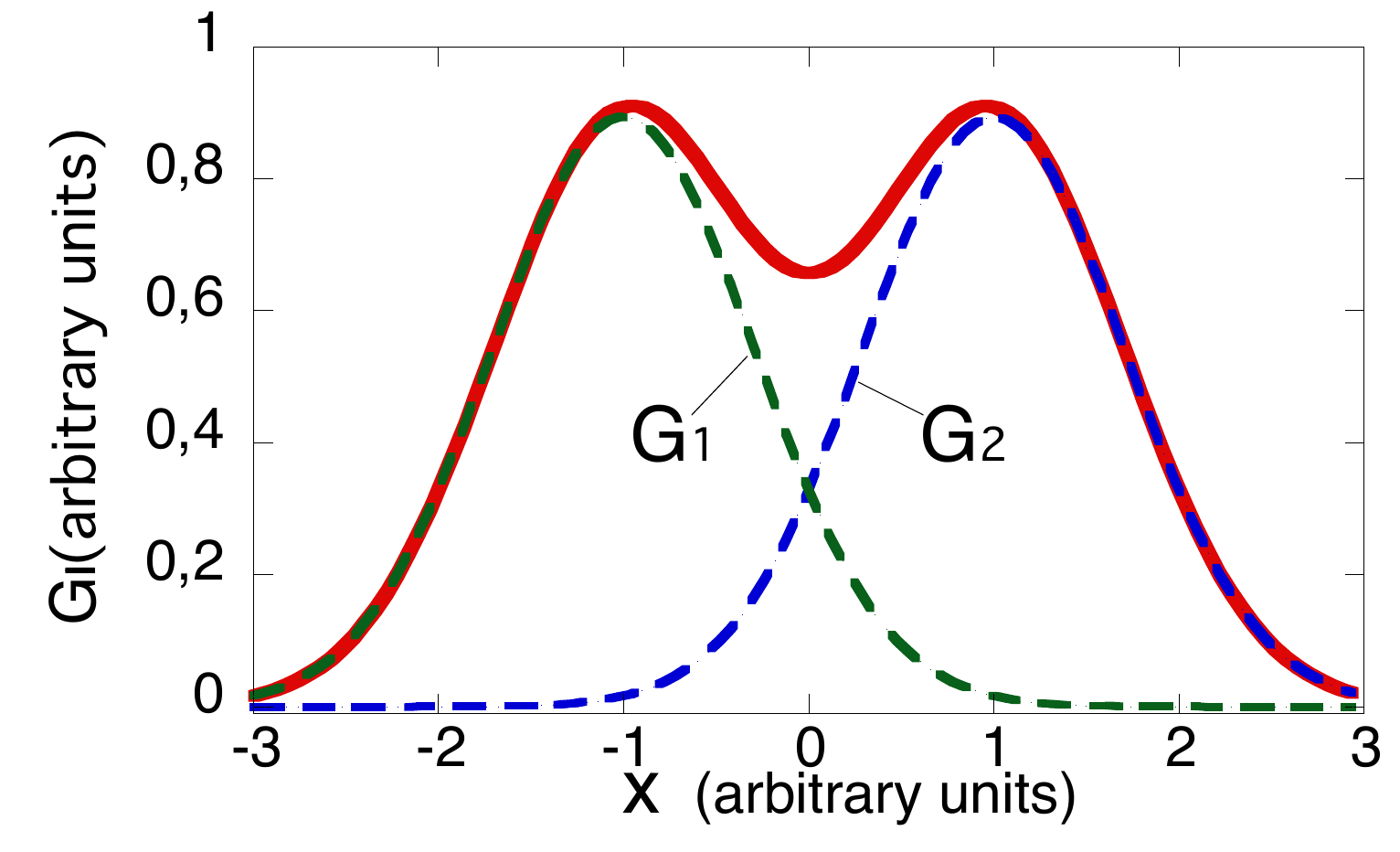}
\label{PROB}
\caption{(colour online) Envelope of a two-hump initial pulse (solid) is a sum of  two Gaussians (dashed). If upon transmission each Gaussian is advanced by a distance $\a$, so will also 
be the entire two-hump envelope.
}
\end{figure},  
\begin{eqnarray}\label{31}
G^0(x)=\sum_{j=1}^J(2/\pi \sigma_j^2)^{1/4} \exp[-(x-a_j)^2/\sigma_j^2]\nonumber \\
\equiv \sum_{j=1}^J G_j(x-a_j).
\end{eqnarray} 
Consider again the model of Section IV.
If all $\sigma_j$ in Eq.(\ref{31}) are chosen sufficiently large for each of the Gaussian pulses to experience an accurate
'superluminal' advancement by, say, $\alpha$, the same advancement will be experienced by their multi-hump sum, 
whose envelope will evolve into (as before, we put $X\equiv x-p_0t$)
\begin{eqnarray}\label{31a}
\sum_{j=1}^J G_j(X-a_j-\alpha)= G^0(X-\a).
\end{eqnarray}
This is an important addition to the argument of Ref.\cite{BUTT1}. The transmitted pulse is carved not from front part of the whole pulse, but from the front parts of each of its constituents Gaussians. This clearly distinguishes the quantum reshaping mechanism from the scissors-based
 reshaping of Sect. II.
 \newline
To avoid answering the question 'How does the system know the pulse is built from many Gaussians?' one may revert to the momentum space. A sufficient condition for an advancement by $\a$ is that the momentum distribution $A(p)$ of the incident pulse should fit into the superoscillatory band shown in Fig. 4. A spacial shift by $a_j$ does not broaden $A(p)$, but multiples it by $\exp(-ia_j p)$. Thus, for any choice of $a_j$, the composite pulse will fit into the band and will be advanced as a whole. Given the similarities in the analysis, the same argument will apply to tunnelling across a potential barrier, including the Hartman case discussed in Sect. XVI.
\newline
This should not be confused with the possibility of transferring information at superluminal speeds.
It is true that one can distinguish between one-hump and two-hump signals before the freely propagation humps arrive at the detector. It is also true that the front tails of all the Gaussians involved are already at the detector, and one only needs a clever way to detect them.
A device effecting 'superluminal' advancement acts as a 'filter', making the detection easier, but that is all.
Thus, a barrier 'processes' each component of a double humped pulse separately, this produces a double humped output, and a naive reshaping argument of Sect. II is clearly wrong.
\newline
Yet it is also easy to demonstrate that Winful's own explanation of the effect \cite{REV3}\cite{WIN1}-\cite{WIN3} is insufficient. The author argued that 'energy is stored in the barrier and then released' \cite{WIN1}
in such a way that 'the output adiabatically follows the input with a delay proportional to the stored energy' \cite{WIN2}. Consider, however, an experiment, similar to the one discussed in Sect. XV, 
where the transmitted Gaussian pulse lies several widths in front of the freely propagating one.
Now modify the initial pulse by chopping off its rear half, but leaving its front half intact.
Let us forget for a moment about wave packet spreading. The convolution formula (\ref{15}) ensures that the effects of chopping will not be felt to the right of $x=p_0t/\mu$, where all constituent envelopes are, as we said, unchanged. Thus, the transmitted wave packet will be a full Gaussian rather than just its front half. The spreading is not a problem for our argument. One only requires (cf. Eq.(\ref{15})) that a freely propagating chopped pulse would not alter significantly its front part by the time we take a snapshot of what is transmitted. This can be ensured by either making the cut smooth, considering a heavy particle with a large $\mu$, or choosing a high barrier and a large $p_0/\mu$, so the pulse will not have enough time to spread. We see then that the effect cannot be explained by the output 'adiabatically following the input' as suggested in \cite{WIN3}.
There is no input and no energy storage corresponding to the missing rear part of the Gaussian, and yet the output has it. A detailed calculation for our model, which shares the same mechanism with tunnelling, can be found in \cite{DS4a}. An experimental observation of a similar effect was reported in \cite{NATUR}, where it was shown that the advanced part of a pulse propagating in a fast-light medium does not carry information about any modifications made to the rear part of initial signal. Had this not been the case, a reliable superluminal communication would have been possible, and Einstein's causality would have been in trouble. Along with other authors we stress that this is not the case. On the other hand, the mechanism proposed in \cite{REV3}\cite{WIN1}-\cite{WIN3} fails to address one of the most intriguing features of the effect and, therefore  cannot 
be a viable candidate for a general explanation of  apparent 'superlumunality' in wave packet propagation.
\newline 
Finally, to achieve 'superluminal' transmission, one only requires certain behaviour of the transmission amplitude, and a suitable choice of the incident wave packet. The former can be achieved in various ways, e.g, through entanglement with a subsequent post-selection, as in the model of Sects. III - X, or by passing across the classically forbidden region, or by other means \cite{MUGA1}, \cite{MUGA2}. In all cases this appears to be a specific effect, unlikely to be reduced to simpler physical explanations. 
All one can say is that it is an interference phenomenon, similar to the one causing the appearance of 'anomalous' values in the 'weak' measurement theory,  in which none of the interfering components 'move faster' than in free motion.
\section { Conclusions and discussion}
In summary, the phenomenon of apparent 'superluminality' can be analysed in the coordinate or in the momentum space. Each approach has its advantages. In the coordinate representation, a potential barrier (or, indeed, any system with a linear relation between the incident and transmitted amplitudes) acts as an effective beamsplitter, recombining weighted copies of the incident pulse with all spacial shifts $x'$, into the transmitted one. For a potential supporting no bound states, there are no positive shifts, i.e., none of the pulses leaving the beamsplitter are advanced, $x'\le 0$. (Note the classical analogy:  a particle cannot be sped up unless it passes over a potential well where its velocity increases). The causal nature of the propagation is, therefore, stated explicitly.

Preparing a particle as a wave packet around a mean momentum $p_0$ and determining the particle's position once it has been transmitted, amounts to conducting a quantum measurement of this shift.
This is hardly surprising, since comparing the positions of a classical particle passing through a potential, and of the one moving in the free space, does just that. Classically, one can also go a step further, and relate this spacial advancement, or delay, to the duration the particle's trajectory spend in the potential. Quantally, such a relation does not exist,
and the speculation about a tunnelling time representing  'actual duration spent in the barrier' \cite{REV} is fruitless.

Quantum measurements obey known quantum rules. The uncertainty principle effectively states that two or  more interfering (virtual) pathways are but a single pathway connecting initial and final states of the system. Real pathways are produced, and the 'which way?' question can be answered, only if an interference is destroyed, e.g., by a measurement. In tunnelling, the situation is similar to the one which arises in Aharonov's  weak measurements. The accuracy to which the shift $x'$ is measured by detecting the transmitted particle in $x$ depends on the wave packet width $\sigma$,  
with the shifts $x-\sigma \lesssim x' \lesssim x+\sigma$ still allowed to interfere. An accurate measurement with a small $\sigma$ always finds $x'\le0$. For an inaccurate 'weak' measurement with a large $\sigma$ the question 'which shift?' becomes meaningless due to the interference.
A question for which there should be no answer provokes a 'silly' answer \cite{DSNEGAT}, \cite{DSNEGAT1}:  a forward  shift by the barrier width $d$, even though the 'beamsplitter' shifts all the components of the transmitted pulse backwards.
This result translated into the language of tunnelling times and taken at the face value, constitutes the 'superluminal paradox'.

 While the coordinate representation is best suited to expose the causal nature of transmission by comparing it to a failed measurement, the momentum space offers a description in terms of 'superoscillations'. Causality, formulated above as the absence of negative shifts, also  requires that the Fourier integral of $T(p)$ contains only exponentials with positive frequencies. In the momentum space the 'paradox' consists in that to advance a pulse by $\a$ one needs $T(p)$ to behave as $\exp(-i\a p)$, while such exponentials are absent from its Fourier integral. There its resolution is the 'superoscialltions' phenomenon studied by Berry and others \cite{SOSC}: the ability of exponentials with non-negative frequencies to mimic, locally, the behaviour of a plane wave with a negative frequency.
If so, a pulse narrow in the momentum space probes this local superoscilltory behaviour of the transmitted amplitude and is advanced, unconcerned about the global analytical properties of $T(p)$.

To conclude, apparent 'superluminality' is, in essence, an interference effect. It cannot be adequately described by a classical analogy such as the naive reshaping of Sect. II, or the energy storage mechanism. 
To our knowledge, the most detailed description of the effect can be achieved in terms of quantum measurement theory, where there exists a language designed to deal with a similar type of interference phenomena.
Finally, the question of which microscopic properties make a particular material behave as barrier for particles or waves is beyond the scope of this paper. Its analysis can, however, be applied to transmission of wave packets of various nature in different types of physical media.
\section{Acknowledgements}
One of us (DS) acknowledges support of the Basque Government Grant No. IT472, MICINN (Ministerio de Ciencia e Innovacion) Grant No. FIS2009- 12773-C02-01 and of the UPV/EHU under the program UFI 11/55.
\section{Appendix. Von Neumann measurements with post-selection}

To emphasise the analogy with Eqs.(\ref{2}) and (\ref{15}), we briefly discuss a von Neumann quantum measurement \cite{vN} of an operator $\ah$ with a continuum spectrum of negative eigenvalues, performed by means of meter with position $x$, coupled linearly in the pointer's momentum.
At $t=0$ the system is prepared in an initial state
\begin{eqnarray}\label{Ap1}
|a\ra=\int_{-\infty}^{0} a(A) |A\ra d A, \q \ah|A\ra=A|A\ra
\end{eqnarray} 
and the pointer in the state
\begin{eqnarray}\label{Ap2}
|\Psi_{I}\ra=\int_{-\infty}^{\infty} G^0(x)|x\ra dx, 
\end{eqnarray}
e.g., a Gaussan (\ref{7}), of  a width $\sigma$.
A coupling
\begin{eqnarray}\label{Ap3}
 \h^{int}=-iT^{-1}\partial_x \hat{A}
\end{eqnarray} 
is switched on briefly for $0\le t\le T$, after which one finds the pointer and the system  in an entangled final state
\begin{eqnarray}\label{Ap4}
\la x |\Psi_{T}\ra=\int_{-\infty}^{0} G^0(x-A)a(A) |A\ra dA.  
\end{eqnarray}
Suppose we also check the state of the system at $t=T$, and collect the statistics of the pointer's final position only if the system is found in a state $|b\ra=\int_{-\infty}^{\infty} b(A) |A\ra d A$. 
In this post-selected ensemble, the probability to find the pointer at $x$, 
is $P^{b\leftarrow a}(x) = |\Psi^{b\leftarrow a}(x)|^2/\int|\Psi^{b\leftarrow a}(x)|^2dx $, where $\Psi^{b\leftarrow a}(x)$ 
is given by a convolution formula similar to Eqs. (\ref{2}) and (\ref{15}),
\begin{eqnarray}\label{Ap5}
\Psi^{b\leftarrow a}(x)=\int_{-\infty}^{0} G^0(x-A)\eta(A) dA, \\
\nonumber
 \eta(A)\equiv b^*(A)a(A).
\end{eqnarray}
In the limit  $\sigma\to 0$ one has a {\it strong} measurement, with the expectation value, 
[$\la x\ra \equiv \int xP^{b\leftarrow a}(x) dx$] 
\begin{eqnarray}\label{Ap6}
\la x\ra=\int_{-\infty}^{0} A |\eta(A)|^2 dA/\int_{-\infty}^{0} |\eta(A)|^2 dA \le 0.
\end{eqnarray}
In the opposite limit, $\sigma\to \infty$, one obtains a {\it weak} measurement \cite{AH1}-\cite{AH3}, \cite{DSNEGAT}, with the 
expectation value 
\begin{eqnarray}\label{Ap7}
\la x\ra=Re \bar{A},\quad \quad \quad \quad  \quad \quad \quad \quad \quad \quad
 \\ \nonumber \bar{A}\equiv  \int_{-\infty}^{0} A \eta(A) dA/\int_{-\infty}^{0} \eta(A) dA
\end{eqnarray}
 and a variance of order of $\sigma$. Unlike the r.h.s of Eq.(\ref{Ap6}), $Re \bar{A}$ is not restricted to the negative semi-axis, and for certain choices of $|a\ra$ and $|b\ra$ may be large positive. Such an $\bar{A}$ is called an {\it anomalous weak value}. Finally, if $Re \bar{A}> \sigma$, so that all the pointer's readings occur outside the spectrum of $A$, an anomalous weak value is {\it sharp}.


\begin{thebibliography}{999}
\bibitem{FOOTSC} Except in the (semi-)classical case, where there exists a direct relation 
between particle's position and the duration its trajectory spends in the region containing potential
\end{thebibliography}
\end{document}